\documentclass[preprint]{aastex631}
\usepackage{amsmath}
\usepackage{amssymb}

\newcommand{\kla}[1]{\left<{#1}\right>}
\newcommand{\boldu}{\boldsymbol{u}}

\newcommand{\boldk}{\boldsymbol{k}}

\newcommand{\boldn}{\boldsymbol{n}}
\newcommand{\boldB}{\boldsymbol{B}}

\begin{document}

\title{On the Interpretation of the Scalings of Density Fluctuations from In-situ Solar Wind Observations: Insights from 3D Turbulence Simulations}

\author[0000-0003-1134-3909]{Senbei Du}
\affiliation{Los Alamos National Laboratory, Los Alamos, NM 87545, USA}

\author[0000-0003-3556-6568]{Hui Li}
\affiliation{Los Alamos National Laboratory, Los Alamos, NM 87545, USA}

\author[0000-0003-3886-0383]{Zhaoming Gan}
\affiliation{New Mexico Consortium, Los Alamos, NM 87544, USA}

\author[0000-0002-4305-6624]{Xiangrong Fu}
\affiliation{New Mexico Consortium, Los Alamos, NM 87544, USA}

\begin{abstract}

Solar wind turbulence is often perceived as weakly compressible and the density fluctuations remain poorly understood both theoretically and observationally.
Compressible magnetohydrodynamic simulations provide useful insights into the nature of density fluctuations.
We discuss a few important effects related to 3D simulations of turbulence and in-situ observations.
The observed quantities such as the power spectrum and variance depend on the angle between the sampling trajectory and the mean magnetic field due to anisotropy of the turbulence. The anisotropy effect is stronger at smaller scales and lower plasma beta.
Additionally, in-situ measurements tend to exhibit a broad range of variations, even though they could be drawn from the same population with the defined averages, so a careful averaging may be needed to reveal the scaling relations between density variations and other turbulence quantities such as turbulent Mach number from observations.

\end{abstract}

\section{Introduction}

Turbulence has been a key subject for space plasma physics for many decades \citep{Coleman1968}. Solar wind turbulence is typically perceived as weakly compressible with dominant Alfv\'enic signatures \citep{Belcher1971} and the compressible component usually contains a small fraction ($\sim 10\%$) of the total fluctuation energy \citep[e.g.][]{howes_apj_2012}.
As a result, observations and theories of solar wind turbulence often focus on the incompressible aspect. The compressive and density fluctuations remain poorly understood.
Perhaps the most well developed theory for compressible solar wind turbulence is the nearly incompressible (NI) theory. The NI theory is applicable in the limit of low turbulent Mach number and low Alfv\'en Mach number, where the compressible magnetohydrodynamic (MHD) system is dominated by the 3D incompressible MHD at high plasma beta $\beta$ (ratio between the thermal and magnetic pressure), and 2D incompressible MHD at moderate to low beta \citep{Zank1992,Zank1993}.
The density fluctuations $\delta \rho$ enter at the second order, which leads to a scaling of $\delta \rho / \rho_0 \propto M_t^2$, where $M_t = \delta v/c_s$ is the turbulent Mach number and $c_s$ is the sound speed.
The origin of density fluctuation in NI thoery is postulated to be the ``pseudo-sound'' process, where the density variation follows the eigenrelation of a sound wave but is produced by incompressible fluid motion instead \citep{Lighthill1952,Montgomery1987}.
Extension of the NI theory in an inhomogeneous background can also produce a linear scaling of $\delta \rho / \rho_0 \propto M_t$ in the solar wind \citep{Bhattacharjee1998,Hunana2010}.
A linear scaling can also be expected in a linear wave description where turbulent fluctuations are decomposed into linear MHD modes \citep{Cho2003}, but analysis based on frequency-wavenumber spectra suggests that propagating waves may not be a good representation of turbulence \citep{Gan2022}.
Observation data have been inconclusive regarding the scaling mostly due to the wide range of variation that tends to exist in the data \citep[e.g.,][]{Matthaeus1991,Adhikari2020}.

Another useful approach to study turbulence is through numerical simulations.
Compressible MHD turbulence simulations have been performed in the past to investigate the density fluctuations.
\citet{Matthaeus1996} find that density fluctuations perpendicular to the mean magnetic field are stronger than those parallel to the mean magnetic field. The level of anisotropy in density fluctuations tends to be between that of longitudinal and transverse velocity fluctuations.
This is in general consistent with the NI theory where most of the density fluctuations are generated by the pseudo-sound mechanism due to 2D fluctuations.
(In the NI theory, the 2D fluctuations of velocity are also thought to be responsible for the decay of density fluctuations \citep{Zank2017}.)
Based on the decomposition into linear modes, \citet{Cho2003} find critical balance-like anisotropy \citep{Goldreich1995} for Alfv\'en and slow modes, while fast modes are found to be approximately isotropic. Recently, \citet{Fu2022} investigate the scaling of density fluctuation by performing simulations with varying amplitudes of the external driving forces that are applied to the MHD momentum and induction equations. A linear scaling of $\delta\rho/\rho_0$ as a function of $M_t$ is found for all values of plasma beta and cross helicity.
The differences between the simulation results and the various theoretical predictions suggest that there is still a lack of fundamental understanding of the scalings of density fluctuations in turbulence.

The most straightforward way of comparing simulations with observations is to extract simulation data from a fixed time step. By doing this, the standard Taylor's hypothesis is assumed, where turbulence is regarded as a composite of spatial structures \citep{Taylor1938}. As we will show in this paper, even when Taylor's hypothesis is valid, there are still nuances that need to be considered when comparisons are attempted.

In this paper, we will first revisit Taylor's hypothesis in Section \ref{sec:taylor}. In Sections \ref{sec:simulation} and \ref{sec:result}, we present our 3D MHD simulations and their results, emphasizing how the anisotropy and sampling will affect interpretation of density scalings. In Section \ref{sec:discussion}, we summarize our results and discuss their implications.

\section{Taylor's hypothesis revisited}\label{sec:taylor}

Taylor's hypothesis \citep{Taylor1938} is a cornerstone for interpreting single-spacecraft in-situ observations of the solar wind.
The basic idea is that temporal lags ($\tau$) in observational data can be directly translated to spatial separations ($l$) based on the bulk velocity of the wind $u_{sw}$, i.e., $l= u_{sw} \tau $.
Essentially, this assumes that the observed turbulence consists of spatial structures that are advected with the flow. Taylor's hypothesis is valid when the characteristic wave velocities are much smaller than the flow velocity so that the temporal variation can be neglected. This is typically the case for solar wind near 1 au where the wind speed is $\sim 400$ km/s and the Alfv\'en speed is $\sim 20$ km/s.
There are situations, especially close to the Sun, where the wind speed is comparable or even smaller than the Alfv\'en speed. In those cases, Taylor's hypothesis will need to be modified to take into account the wave propagation.
Such an approach is adopted by \citet{Zank2022} as they analyze the \textit{Parker Solar Probe} data near the Alfv\'en critical point where the Alfv\'en speed equals the wind speed.
For the present work, we consider only cases where Taylor's hypothesis is applicable.

In terms of spectral analysis, Taylor's hypothesis can be expressed in a more precise manner, as shown by \citet{Fredricks1976}. (Actually, the work by \citet{Fredricks1976} concerns more general cases that include wave propagation.) Assuming the 3D wavenumber spectrum is $P(\boldk)$, the observed frequency spectrum is given by
\begin{equation}\label{eq:Pomega}
  P(\omega) = \int d^3 k P(\boldk) \delta(\omega - \boldk\cdot \boldu_{sw}).
\end{equation}
The relation can be shown by noting the definition of the 3D power spectrum
\begin{equation}
  P(\boldk) = (2\pi)^3 \int d^3 \Delta x R_x(\boldsymbol{l}) e^{-i\boldk\cdot\boldsymbol{l}},
\end{equation}
which is the Fourier transform of the correlation function $R_x(\boldsymbol{l})$ (assuming homogeneous turbulence). For single-spacecraft measurements, the temporal correlation function can be expressed with the spatial correlation function, and thus the 3D power spectrum, as
\begin{equation}\label{eq:Rt}
  R_t(\tau) = R_x(\boldu_{sw} \tau) = \int d^3 k P(\boldk) e^{i\boldk\cdot\boldu_{sw}\tau}.
\end{equation}
Then the frequency spectrum \eqref{eq:Pomega} can be obtained by taking the Fourier transform of Equation \eqref{eq:Rt} and using the properties of the Dirac delta function.
Recent works by \citet{Bourouaine2019} and \citet{Perez2021} suggest that even when Taylor's hypothesis is applicable, the velocity fluctuations can introduce a random sweeping effect that broadens the observed frequency spectrum. This will not be considered in the present study because we are currently considering the wavenumber spectrum from a single time frame of simulations, and the sweeping effect is not present in this situation.

For anisotropic turbulence, it is evident that the observed frequency spectrum depends on the direction of $\boldu_{sw}$.
In magnetized plasmas such as the solar wind, the turbulence is typically approximately gyrotropic with respect to the background magnetic field, so that the power spectrum can be written as $P_{2D}(k_{\parallel}, k_{\perp})$. This is referred to as the 2D reduced spectrum, and it relates to the 3D spectrum as $\int d^3 k P_{3D}(\boldk) = \int dk_{\parallel} dk_{\perp} P_{2D}(k_{\parallel}, k_{\perp})$.
If one considers the 1D spectrum sampled along a direction of $\hat{\boldn}$ that makes an angle $\theta$ from the background magnetic field, it can be shown that \citep{Forman2011}
\begin{equation}\label{eq:P1d}
  P_{1D}(k, \theta) = \int P_{2D}(k_{\parallel}, k_{\perp})\delta(k_{\parallel}\cos\theta + k_{\perp}\sin\theta - k) dk_{\parallel} dk_{\perp}.
\end{equation}
We note that $k$ is regarded as a parameter in the integration and does not necessarily satisfy the relation $k_{\parallel}^2 + k_{\perp}^2 = k^2$.

Another common turbulence measurement is the variance, which represents the intensity of turbulent fluctuations. Strictly speaking, the variance of a quantity $X$ is defined by the ensemble average $Var(X) = \kla{(X - \kla{X})^2}$.
In practice, the ensemble average is replaced by spatial or temporal average.
The ``true'' variance in a 3D simulation can be estimated simply by spatial averaging $\delta X^2 = \overline{(X - \bar{X})^2}$.
For spacecraft observations, the variance is estimated by temporal averaging.
The variance can be calculated by integrating the power spectrum, i.e., $Var(X) = \int d^3k P_X(\boldk)$. In principle, the spacecraft-measured variance is independent of the sampling direction, unlike the spectrum, since the variance is the spectrum integrated over all frequencies or wavenumbers. A nuance is that different sample lengths may be needed along different sampling directions to ensure that the variance is independent of the sampling angle. For convenience, we used a fixed sample length when considering the variance in this paper, and this is also the case for most previous observations.
We will show in our results that the variance does depend on the sampling direction in this case.

The effect of sampling arises when the ensemble average in the 3D spatial correlation function is replaced by the line average along the observer's path, which introduces variations to the observed temporal correlation function, and thus the frequency spectrum. This is likely a finite-sample effect as a 1D sample contains only limited statistical information of the full 3D turbulence.

To summarize, when interpreting spacecraft measurements (which is typically taken as a path through a turbulent volume), we need to be mindful about the intrinsic anisotropic nature of turbulence and the sampling effects. In the next sections, we will demonstrate their impact in detail.

\section{MHD simulation of turbulence}\label{sec:simulation}

We present 3D compressible MHD simulations using the \textit{Athena++} code \citep{Stone2020}.
A background magnetic field $\boldB_0$ is applied along $x$-direction. Periodic boundary conditions are used for all three directions.
Continuous driving forces that follow an Orstein-Ulenbeck process are applied to both velocity and magnetic field, following \citet{Gan2022}.

An isothermal equation of state is used. Most of the results are from a simulation with plasma beta (ratio between thermal pressure and magnetic pressure) $\beta = 1$, though comparisons with a $\beta = 0.2$ run is also made.
The turbulent Mach number $M_t = \delta v/C_s$ at the end of the simulation is $\simeq 0.47$.
To save computational resources, the simulations use an elongated box $L_x = 6\pi$, $L_y = L_z = 2\pi$, and the number of cells ($256 \times 512 \times 512$) is reduced by half in the $x$-direction compared to the other two directions. An elongated box is common for magnetized turbulence simulations \citep[e.g.,][]{Beresnyak2009}.
Energy injection is restricted to large-scale modes with wavelengths at least half the simulation box, i.e., $|\boldk| \leq 2$, and the wavenumber is defined such that the minimum wavenumbers are $k_{\perp,min} = 2\pi/L_y = 1$ and $k_{\parallel,min} = 2\pi/L_x = 1/3$.

\section{Results}\label{sec:result}
\subsection{Anisotropic power spectrum from MHD simulations}

\begin{figure}[!ht]
\centering
\includegraphics[width=0.5\linewidth]{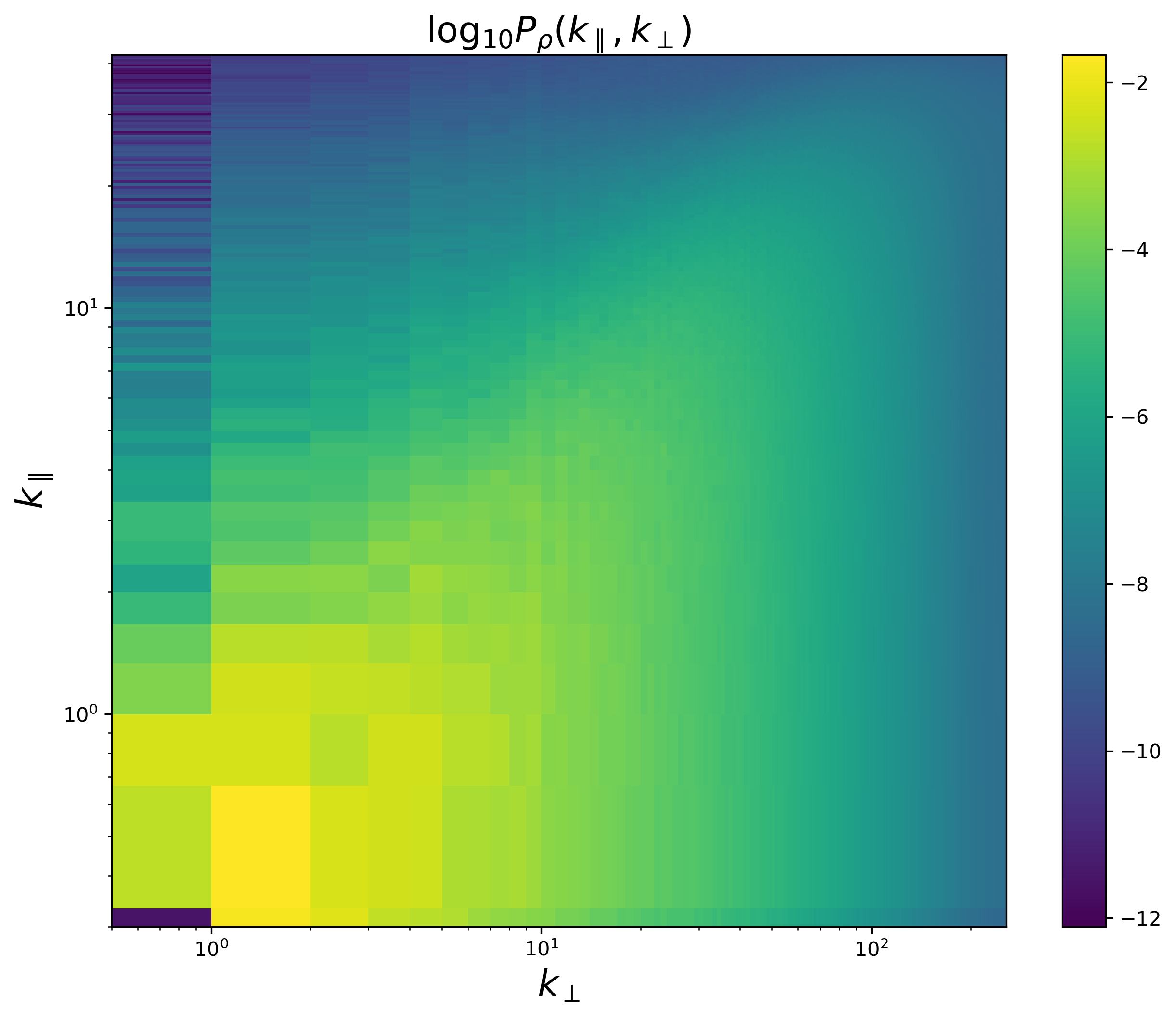}%
\includegraphics[width=0.5\linewidth]{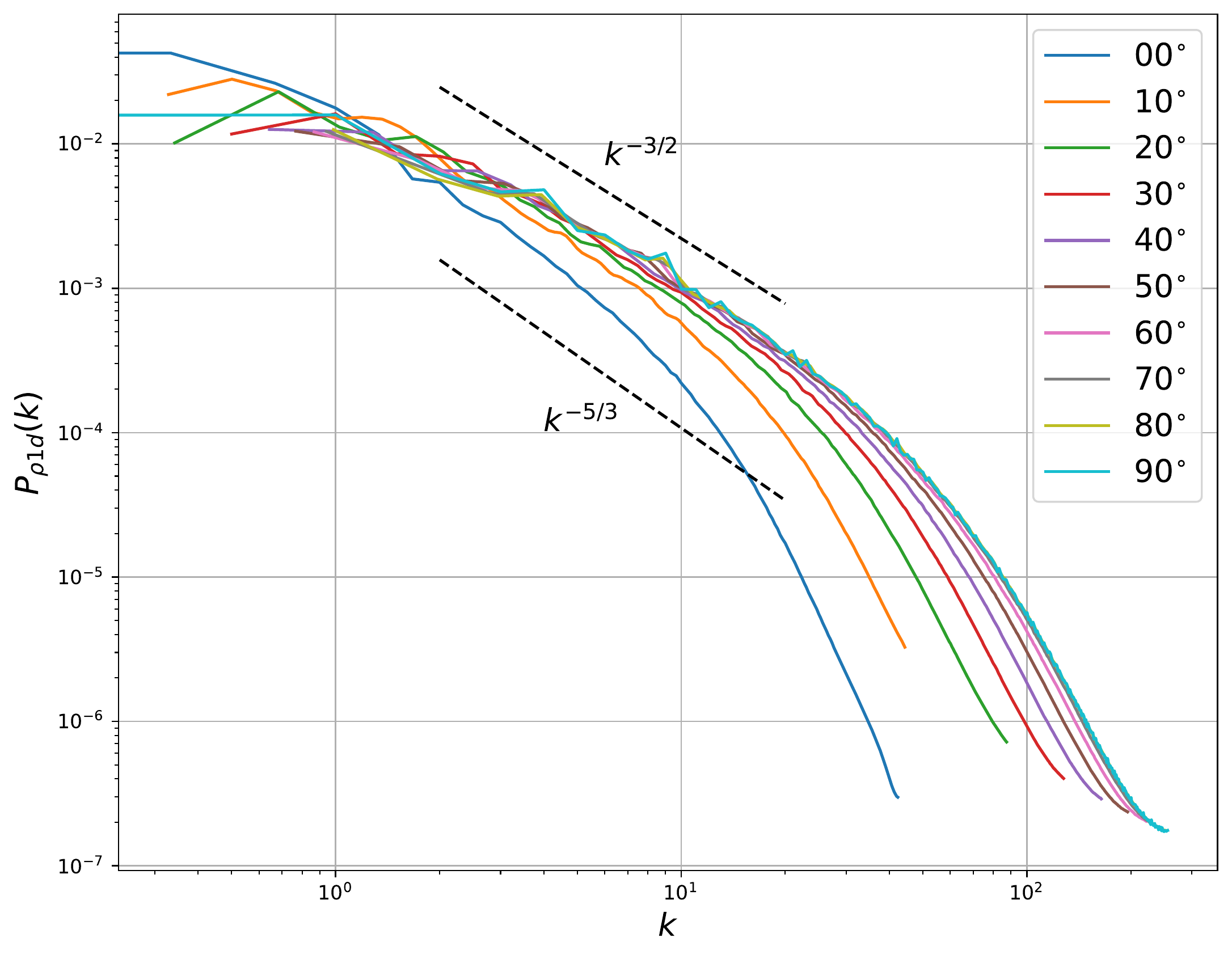}
\caption{Left panel: the 2D power spectrum of density $P_{\rho}(k_{\parallel}, k_{\perp})$ in log scale from the simulation. Right panel: 1D power spectra of density $P_{\rho1d}(k)$ at different sampling angles (defined with respect to the initial background $\boldB_0$).}\label{fig:spectrum-angle}
\end{figure}

To emulate single-spacecraft observations, we calculate the 1D power spectrum of density fluctuations from the simulation using Equation \eqref{eq:P1d}. The 2D spectrum in $k_{\parallel}-k_{\perp}$ space is constructed first, as shown in the left panel of Figure \ref{fig:spectrum-angle}. 
The 1D power spectra at different sampling angles are shown in the right panel of Figure \ref{fig:spectrum-angle}. The 1D spectra clearly exhibit angle-dependent behavior, as the spectral power tends to increase with the sampling angle $\theta$ at a fixed wavenumber.
There is a noticeable change of the spectrum from $\theta=0^{\circ}$ to $\theta=30^{\circ}$, but it remains nearly unchanged at larger angles.
There also appears to be an angle-dependence in the spectral index, as the spectrum is steeper at $\theta = 0^\circ$ compared to the other angles. As a reference, $k^{-3/2}$ and $k^{-5/3}$ power laws are plotted as dashed lines.

The angle-dependent power spectrum is consistent with solar wind observations.
For example, \citet{Horbury2008} shows that the magnetic spectrum has a smaller amplitude and is steeper at smaller angle $\theta$. However, they report that the index changes between -5/3 and -2 at varying angles, which is steeper than the spectra shown here.
It is somewhat surprising that the spectral index anisotropy is present in our analysis although we use the Fourier analysis that do not involve the local mean magnetic field \citep[e.g.,][]{Oughton2020,Yang2021}.
The anisotropic power spectrum may be explained by the recent development in the NI turbulence model \citep{Zank2020}, though a comprehensive theory on density power spectrum is currently lacking.
The density fluctuation spectrum has also been measured near the Alfv\'en critical point \citep{Zank2022}, where it is argued that the density spectrum roughly follows that of 2D incompressible fluctuations.

\subsection{Effects of sampling}

\begin{figure}[!ht]
\centering
\includegraphics[width=0.49\linewidth]{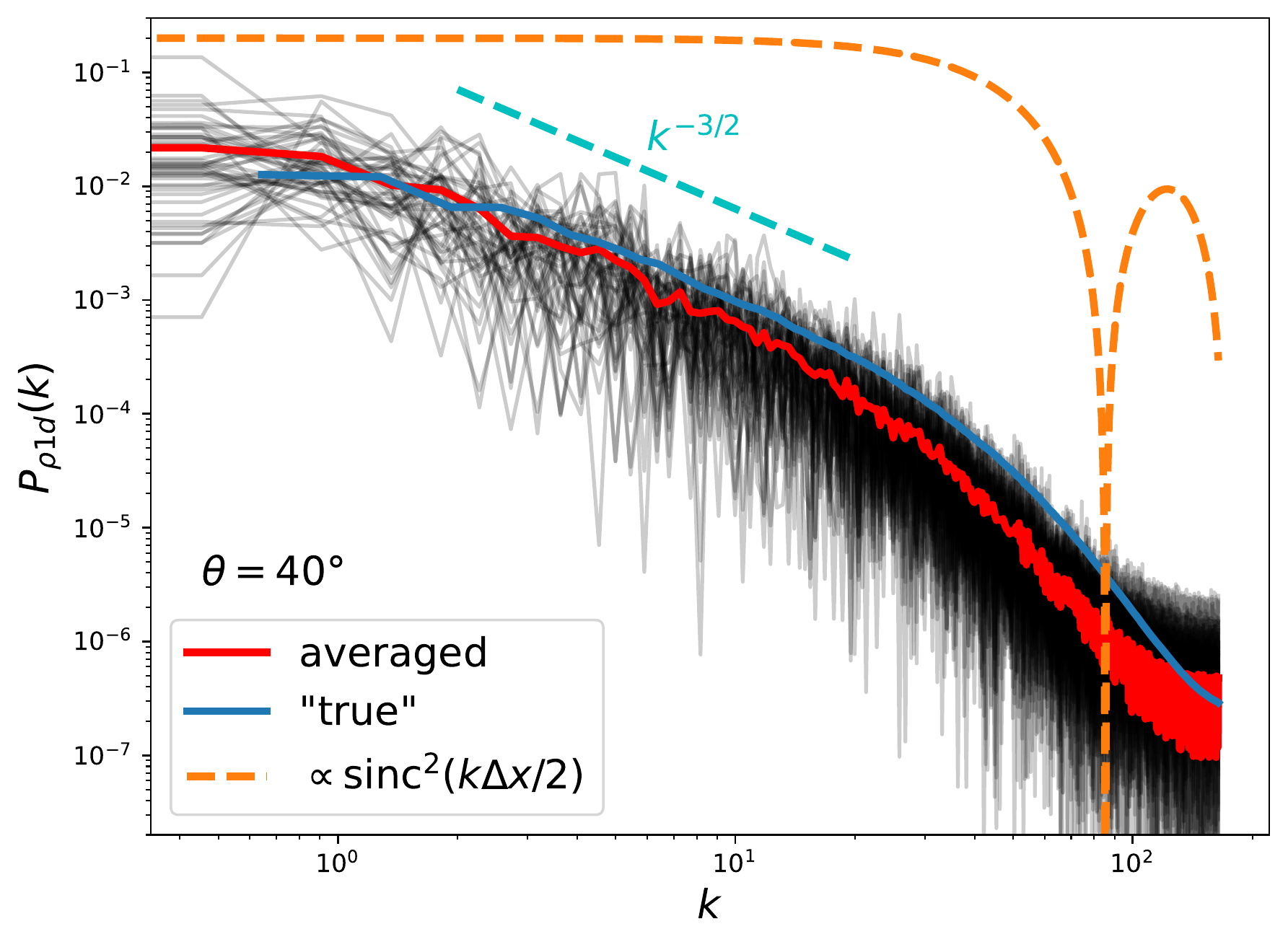}\\
\includegraphics[width=0.49\linewidth]{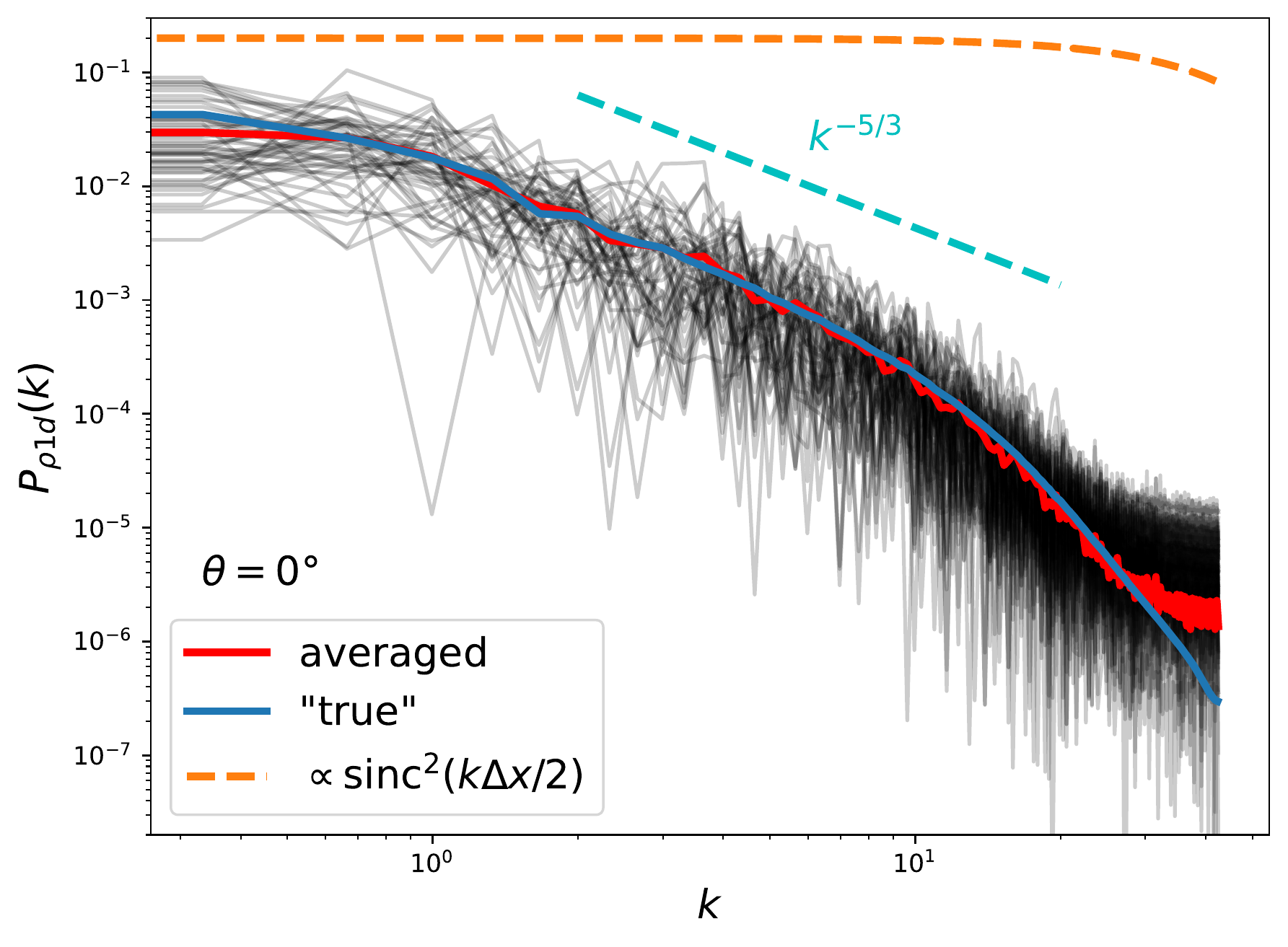}\\
\includegraphics[width=0.49\linewidth]{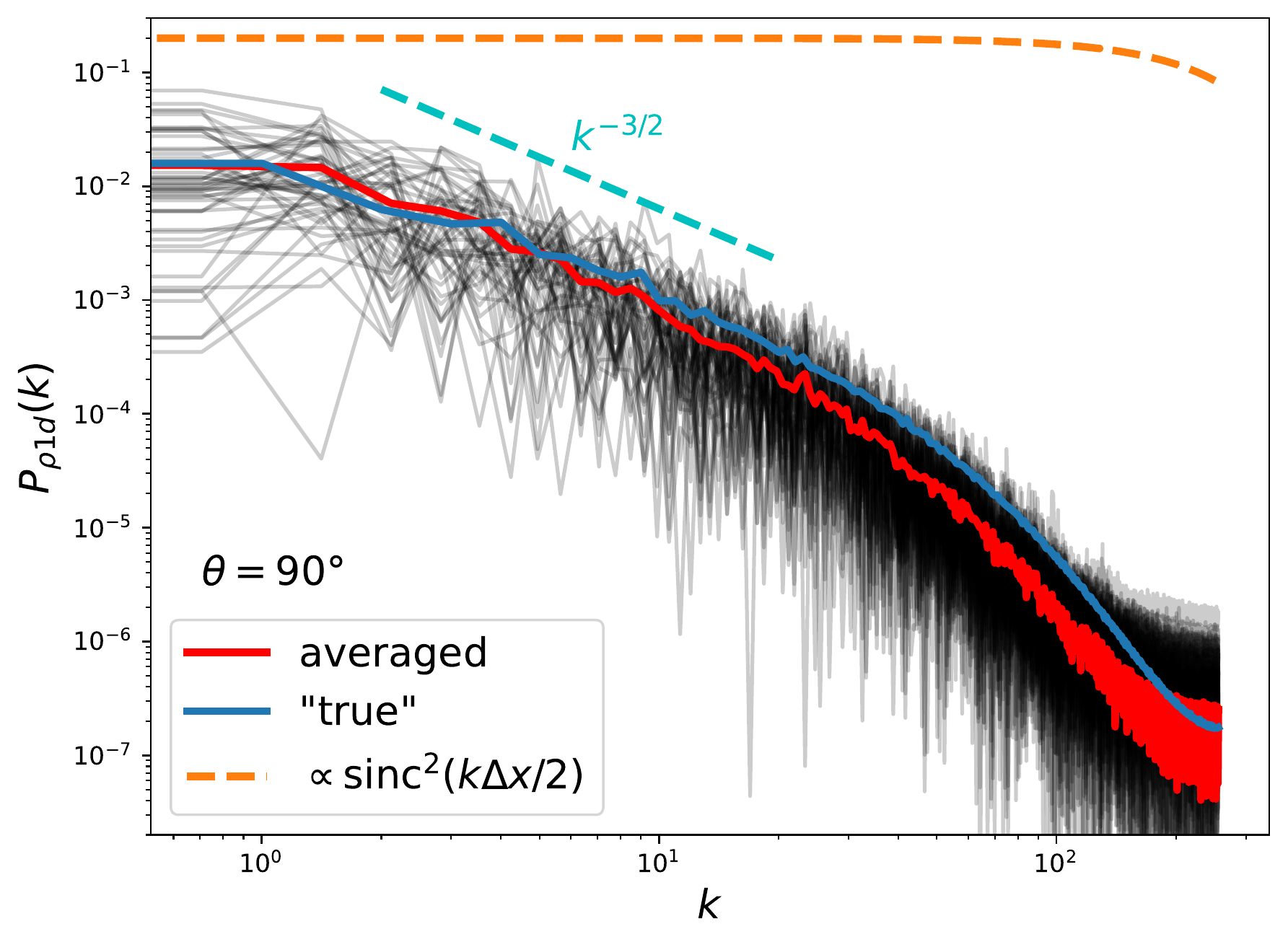}
\caption{Comparison of 1D power spectrum of density calculated along virtual spacecraft samples and 1D reduced spectrum at the same sampling angle of $\theta = 40^\circ$ (top panel), $0^\circ$ (middle panel), and $90^\circ$ (bottom panel). The response function of linear interpolation $\mathrm{sinc}(k\Delta x/2)$ is plotted as a reference. The $k^{-3/2}$ and $k^{-5/3}$ lines are plotted as well.}\label{fig:p-sample}
\end{figure}

The effects of sampling can be assessed by using virtual spacecraft paths to sample the simulation domain. To separate the sampling effects from the anisotropy effects, we select random 1D samples after fixing the angle $\theta$ between the trajectory and the initial background magnetic field ($x$ direction in our simulation). More specifically, for each sample, we first select a random starting point in the simulation domain and then a random azimuthal angle $\Phi$. The length and resolution of the sample is chosen so that the desired wavenumber range is covered. The top panel of Figure \ref{fig:p-sample} shows 50 sample spectra with $\theta = 40^\circ$ as grey lines. The arithmetic average of these spectra is plotted as the red line. Wild fluctuations are present among the sample spectra, and it is interesting that the averaged spectrum does not seem to converge perfectly to the 1D spectrum reduced by Equation \ref{eq:P1d}, shown by the blue line.
Two special cases, $\theta=0^{\circ}$ and $\theta=90^{\circ}$, are shown in the bottom two panels of Figure \ref{fig:p-sample} as a comparison.
A reason for the difference between the reduced spectrum and averaged sample spectrum is suggested by \citet{Bourouaine2020} and \citet{Perez2021}. In general, the reduced spectrum is integrated over a conical surface around the $k_{\perp}=0$ axis in the 3D $k$-space, as suggested by the delta function in Equation \eqref{eq:P1d}, while each sample spectrum corresponds to an integration over a particular plane, as explained in details by \citet{Bourouaine2020}.
For the parallel spectrum, the averaged spectrum should converge to the reduced spectrum, both corresponding to integration over planes perpendicular to the $k_{\perp} = 0$ axis. The middle panel of Figure \ref{fig:p-sample} shows that the reduced and averaged parallel spectra are indeed close to each other.
They show that, for a strict power-law spectrum, the difference between the reduced and the sampled spectrum is a constant factor close to unity.
The difference will be more significant if the intrinsic spectrum is not a single power-law.
Another possible cause for the difference is that 1D sampling requires interpolating the simulation data defined at cell centers to the sample trajectories. The interpolation procedure introduces a wavenumber-dependent response function, which may distort the sample spectrum.
In our case, the linear interpolation is used, whose response function is $F(k) = \mathrm{sinc}^2(k\Delta x/2)$ where $\mathrm{sinc}(x) = \sin(x)/x$ and $\Delta x$ is the grid size.
The function $F(k)$ (multiplied by a constant factor) is plotted in the Figure \ref{fig:p-sample} as the orange dashed lines. For $\theta \ne 90^{\circ}$, we set $\Delta x$ to be $L_x / N_x$, the cell size in $x$ direction; and for $\theta = 90^{\circ}$, we set $\Delta x$ to be the cell size in the $y$ or $z$ direction.
We note that $F(k)$ is a decreasing function of $k$ in the main lobe $k < 2\pi/\Delta x$, which may account for the slightly steeper averaged sample spectrum (red) than the reduced spectrum (blue).
In actual spacecraft observations where the interpolation issue does not exist, we would expect that the observed sample spectra to simply fluctuate around the reduced turbulence spectrum.

\begin{figure}[!ht]
\centering
\includegraphics[width=0.5\linewidth]{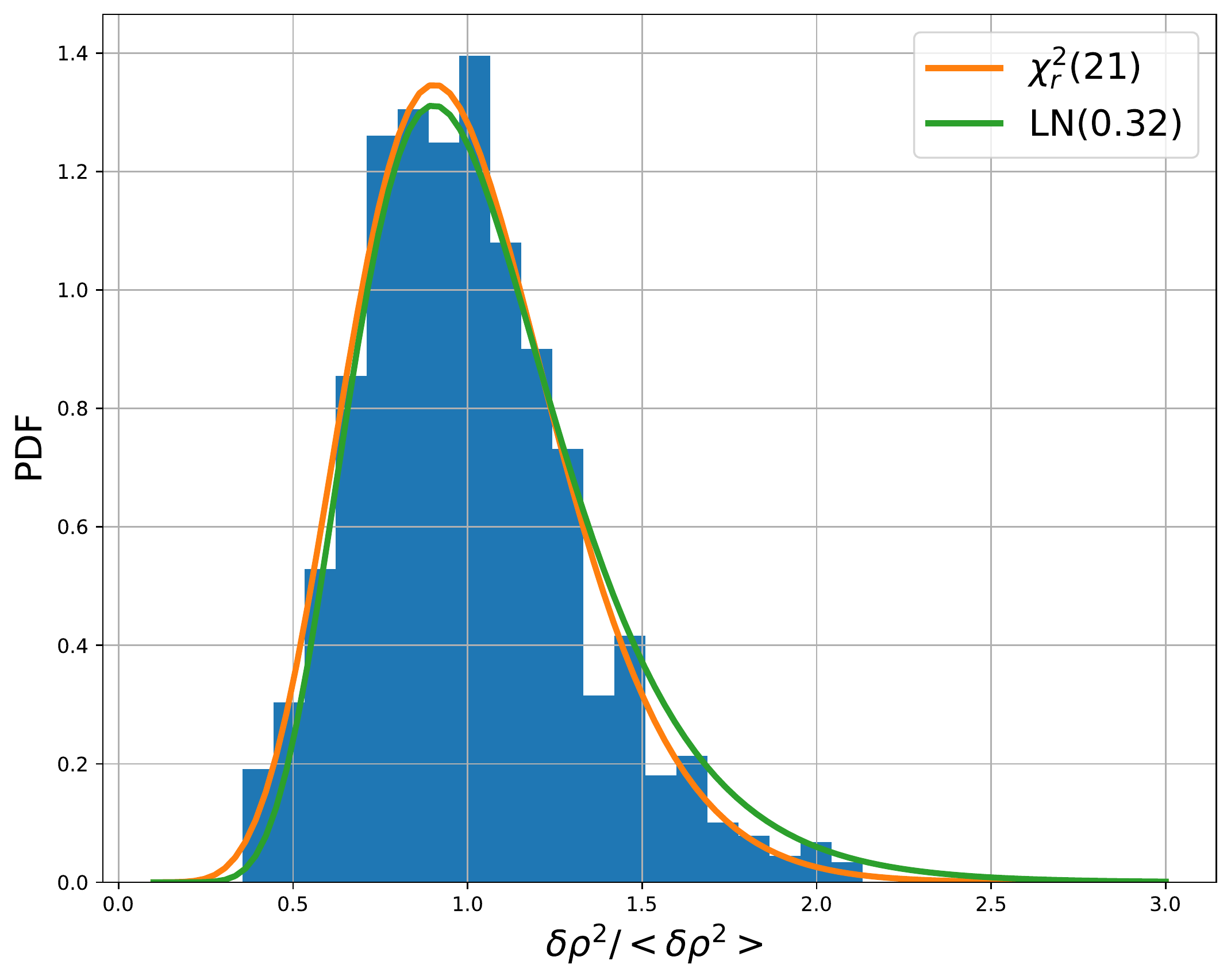}
\caption{Histogram of the normalized density variance, and the comparison with reduced $\chi^2$ and log-normal distributions. The degree of freedom of the reduced $\chi^2$ is set as 21 and the standard deviation of the log-normal is set as 0.32.}\label{fig:pdf}
\end{figure}

A natural question to ask is what types of distribution these sample spectra follow. To simplify the problem, we consider the distribution of variance instead while using a fixed range of wavenumbers.
Figure \ref{fig:pdf} shows the histogram of the density variance among 1000 random samples at a fixed angle $\theta = 40^\circ$.
The variance is calculated by integrating the 1D power spectrum over all wavenumbers above $k = 2$. The variance is then normalized to the mean value of variance $<\delta \rho^2>$ (averaged over all samples).
The PDF of the normalized variance has a shape similar to that of a reduced $\chi^2$ or log-normal distribution.
The reduced $\chi^2$ is a natural choice for the distribution if the density fluctuation can be assumed to be approximately Gaussian, since the variance would then be a sum of the squared Gaussian distributed variables. Though it should be noted that the degree of freedom that is consistent with the distribution (21 in this case) has to be much lower than the number of data points ($\sim 700$) in each sample due to the strong correlation in turbulence signals.
Given suitable parameters, the shape of the log-normal PDF is similar to reduced $\chi^2$, and both distribution is generally consistent with simulation data.

\subsection{Scaling of density fluctuation}

\begin{figure}[!ht]
\centering
\includegraphics[width=0.5\linewidth]{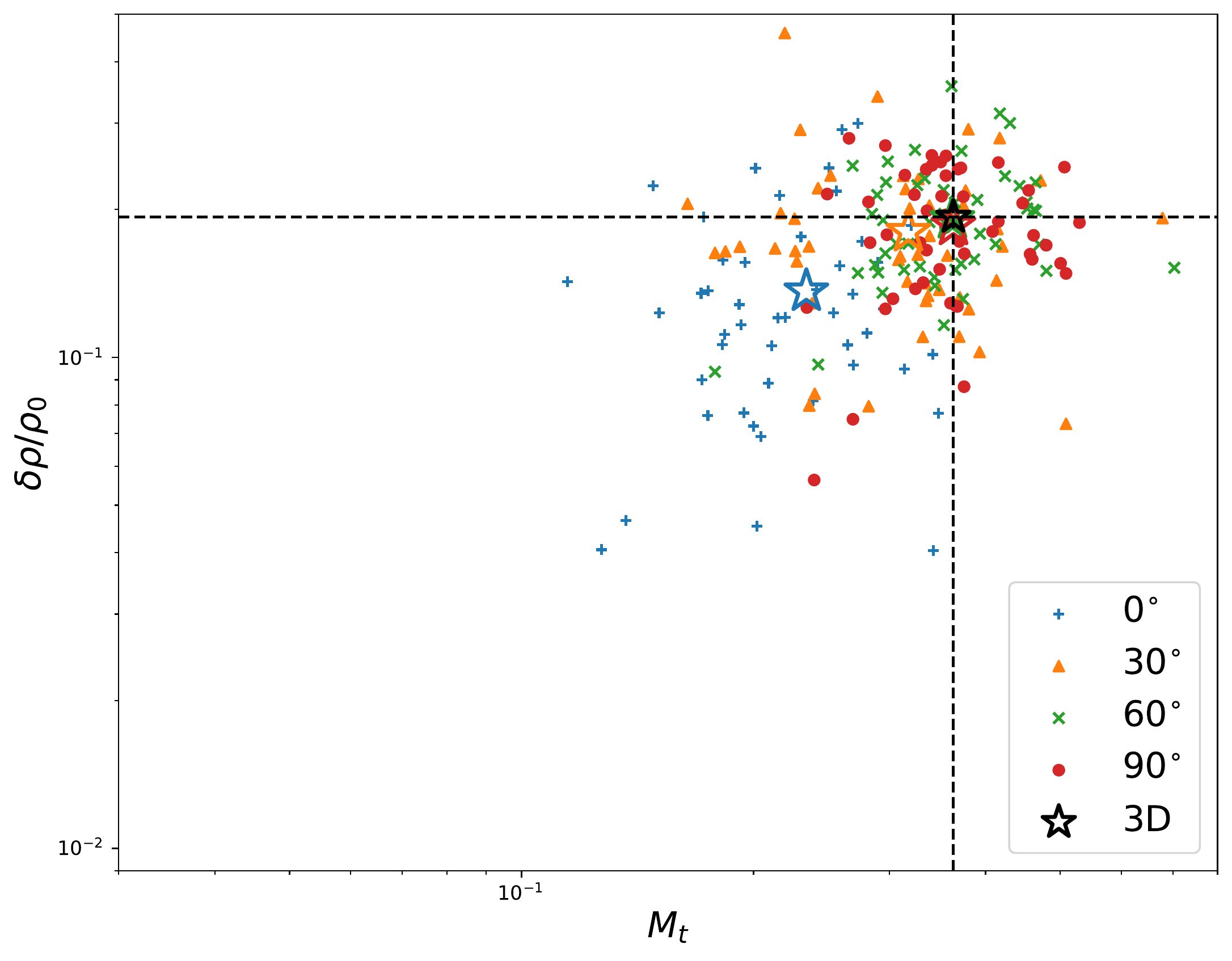}%
\includegraphics[width=0.5\linewidth]{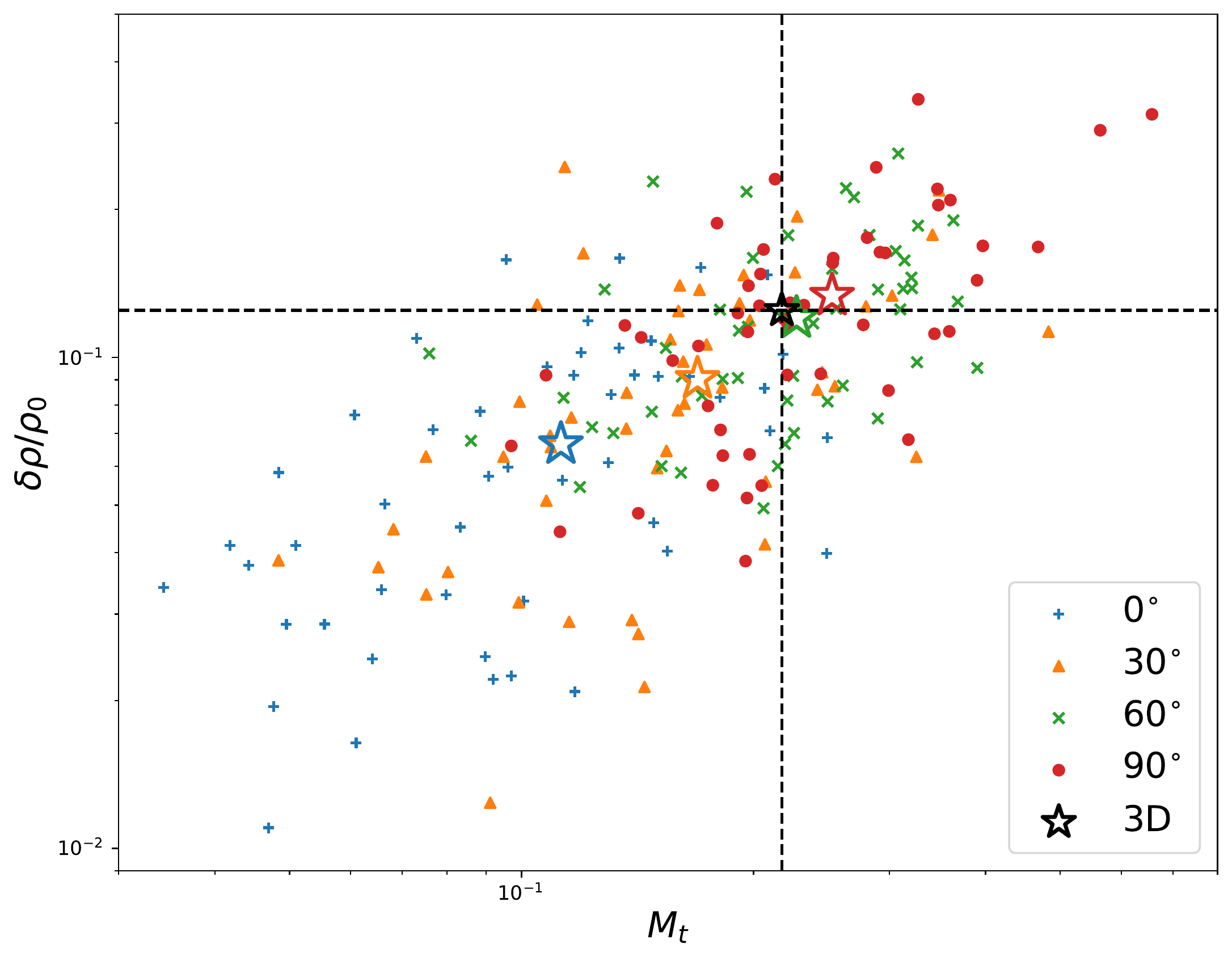}
\caption{Scaling relation between $\delta \rho / \rho_0$ and $M_t$. Different colored symbols represent 1D samples at different angles and the colored stars are the arithmetic mean with the same colored samples. The 3D averaged values with high-pass filters are denoted by the black stars and dashed lines. The left panel uses sample length $\pi$ and the right panel uses sample length $\pi/4$ for all angles.}\label{fig:scaling}
\end{figure}

Here, we investigate how the analysis of the scaling of $\delta \rho / \rho_0$ vs.\ $M_t$ is affected by the anisotropy and sampling.
Figure \ref{fig:scaling} shows an example of the scaling constructed from our simulation. While averaging over the entire 3D domain yields a single number for the density fluctuation and turbulent Mach number as indicated by the black stars and dashed lines, the figure shows that 1D samples introduce significant scattering for both quantities. Here, the sample length on the left panel is fixed at a length of $\pi$, which corresponds to the smallest injection scale ($k=2$). This is compared with the right panel, where the sample length is $\pi/4$ ($k=8$), well into the inertial range.
In each panel, the sample length is the same for all angles although the simulation box is elongated.
Both the anisotropy and the sampling effects are shown in the figure: the different colored symbols represent samples with different angles and the scattering within the same colored symbols is due to sampling effects. Since all the scattered points are from a single simulation, they will obscure the physical scaling relation between $\delta \rho/\rho_0$ and $M_t$.
The comparison of the two panels demonstrates that the anisotropy effect is stronger at a smaller scale, as the different colored stars have a larger variation on the right panel. This is expected since the anisotropic is scale dependent \citep{Goldreich1995,Oughton2020}, which can also be seen in the 1D spectra in Figure \ref{fig:spectrum-angle}.
We note that the ``3D'' values are calculated with high-pass filters, i.e., by integrating the 3D power spectrum over the range $k \ge 2$ (left panel) and $k \ge 8$ (right panel), corresponding to the different sample length.

It is alarming to see that these scattered points already seem to produce an artificial scaling relationship between the density variations and the turbulent Mach number. This is particularly visible in the right panel when shortened 1D samples are used (corresponding to shorter intervals when using observations.)

The effects of anisotropy and sampling are usually overlooked in previous observational studies of the scaling relation. Observations such as by \citet{Matthaeus1991} and \citet{Adhikari2020} are inconclusive because of the big variation among data points. The two effects described here may provide a partial explanation for the observed variation. In addition, the variation of plasma parameters (plasma beta, cross-helicity and adiabatic index) can also play a role.
These effects are probably not a big issue for simulation studies \citep[e.g.,][]{Fu2022}, since the calculation there is based on the full 3D data.

\begin{figure}[!ht]
\centering
\includegraphics[width=1.0\linewidth]{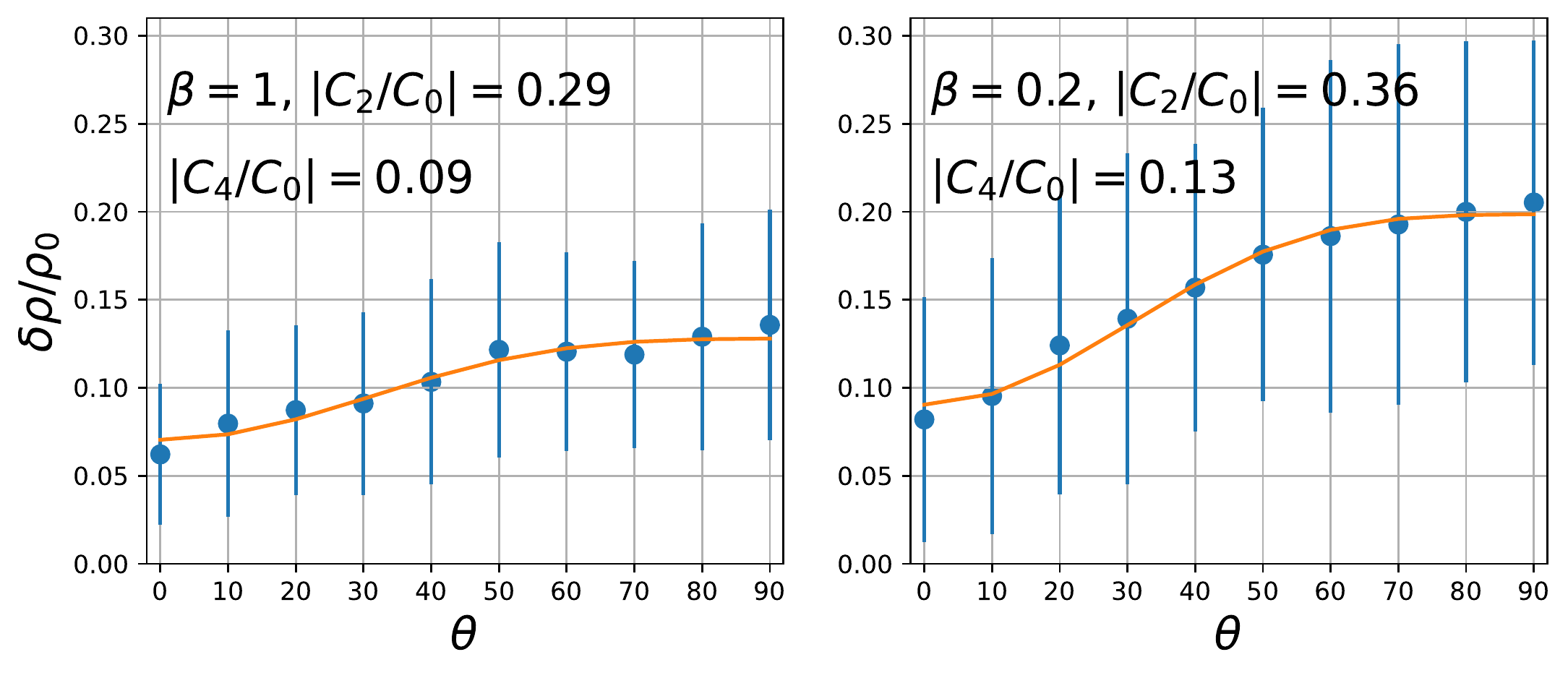}
\caption{The angle-dependent density fluctuation $\delta\rho / \rho_0$ in two simulations with $\beta = 1$ (left panel) and $\beta = 0.2$ (right panel). The 4th-order Legendre polynomial fits are shown as the orange curves. The ratio of Legendre polynomial coefficients $|C_2/C_0|$ and $|C_4/C_0|$ are listed in the figure as a proxy for anisotropy.}\label{fig:beta}
\end{figure}

We also investigate the plasma beta dependence of the anisotropy. Figure \ref{fig:beta} shows the angle-dependent density fluctuation $\delta\rho / \rho_0$ in two simulations with $\beta = 1$ (left panel) and $\beta = 0.2$ (right panel). 100 random samples of length $\pi/4$ at each angle are used for the calculation and the error bars represent the standard deviation of the samples.
To quantify the anisotropy, we employ a 4th-order Legendre polynomial fit for $\delta\rho / \rho_0$ as a function of $\mu = \cos(\theta)$, i.e., 
\begin{equation}
\frac{\delta \rho}{\rho_0}(\mu) = \sum_{i=0,2,4} C_i P_i(\mu),
\end{equation}
where $P_i$ is the $i$-th order Legendre polynomial and $C_i$ is the corresponding coefficient. The fit is shown by the orange curves in the figure. The odd-order coefficients are zero since we assume $\delta\rho(\mu) =\delta\rho(-\mu) $. The ratios between the second and zeroth coefficients $|C_2 / C_0|$ and between fourth and zeroth coefficients $|C_4 / C_0|$ are a proxy for the level of anisotropy.
Our results show that the low-beta simulation has larger $|C_2 / C_0|$ and $|C_4 / C_0|$, indicating a stronger anisotropy therein. This is similar to the conclusion of \citet{Matthaeus1996}.
The apparently opposite conclusion of stronger anisotropy at higher beta is reached by \citet{Cho2003}. This is because the Alfv\'en Mach number $M_A = \delta v / V_A$ are kept constant in their simulations instead of the turbulent Mach number $M_t$, so that the low-beta simulations are supersonic and susceptible to shock formation while high-beta simulations remain subsonic. In contrast, our simulations are all in the subsonic regime.

\section{Discussions and conclusions}\label{sec:discussion}

To summarize, our analysis identifies the following  two effects that will affect the interpretation of observations of solar wind turbulence.

1. Given the intrinsic anisotropic nature of the fluctuations in the solar wind, turbulence measurements such as the power spectrum and variance are angle dependent, i.e., the measured quantity along a spacecraft trajectory depends on the sampling angle between the trajectory (or equivalently, solar wind velocity in the spacecraft frame) and the background magnetic field. 

2. Turbulence measurements can exhibit a wide range of variation among different samples. This is true even if the sampling angle with respect to the background magnetic field is fixed. For any in-situ solar wind turbulence observations, this means that measurements along a single or a small number of sampling intervals are usually not sufficient to draw conclusions about the underlying turbulence.

3. Both the angle dependence and sampling effects can cause artificial trends in scalings such as the density variations versus the turbulent Mach number, as emphasized in Figure \ref{fig:scaling} and \ref{fig:beta}.

To some degree, these issues are known. For example, regarding the first point, \citet{Bieber1996} utilized the angle dependence of inertial-range power spectrum to deduce the dominance of 2D over slab fluctuations; and \citet{Horbury2008} demonstrated that the power and spectral index for magnetic fluctuations are angle dependent. Regarding the second point, the purpose of the commonly used Welch method for estimating PSD is exactly to reduce the variation due to sampling by taking averages over stationary intervals. Decades-long statistical surveys have also been frequently conducted in part to reduce the effects of sampling.

Furthermore, we show that the anisotropy of density fluctuation is scale and beta dependent as stronger anisotropy is found at smaller scales and lower beta.
For in-situ solar wind observations, the effect of anisotropy can be taken into account by considering binning data intervals based on the angle $\theta$, which has led to many fruitful results in the past \citep[e.g.,][]{Bieber1996,Horbury2008}. The effect of sampling can be intuitively reduced by averaging according to the central limit theorem.
The anisotropy of density fluctuation has been noted in simulations by \citet{Matthaeus1996}, but it has not been investigated in details in solar wind data, and this will be done in another publication.
Finally, we show that the two effects can introduce an artificial scaling relation between density fluctuation and turbulent Mach number, which will affect how observational results should be interpreted.
For future observations, we suggest that an averaging procedure should be developed so that the true scaling relation can be inferred.

\begin{acknowledgements}
S. Du and H. Li acknowledge the support by DOE OFES program and LANL/LDRD program. Z. Gan and X. Fu are supported by NASA under Award No. 80NSSC20K0377 and 80NSSC23K0101.
Useful discussions with R.\ Chhiber, M.\ Cuesta, P.\ Kilian, W.\ Matthaeus, J.\ Perez, and J.\ Steinberg are gratefully acknowledged.
This research used resources provided by the Los Alamos National Laboratory Institutional Computing Program, which is supported by the U.S. Department of Energy National Nuclear Security Administration under Contract No.\ 89233218CNA000001.
\end{acknowledgements}

\bibliographystyle{aasjournal}
\bibliography{angle}

\end{document}